\documentclass[preprint,12pt,authoryear]{elsarticle}

\usepackage{graphics}
\usepackage{graphicx}
\usepackage{epsfig}

\usepackage{amssymb}

\usepackage{cite}

\def\arcsec{$^{\prime\prime}$}
\journal{New Astronomy}

\begin{document}

\begin{frontmatter}



\title{Homologous prominence non-radial eruptions: A case study}

\author[label1]{P. Duchlev}
\author[label1]{K. Koleva}
\ead{koleva@astro.bas.bg}
\author[label2]{M.S. Madjarska}
\author[label1]{M. Dechev, }

\address[label1]{Institute of Astronomy and National Astronomical Observatory,\\
 Bulgarian Academy of Sciences, 72 Tsarigradsko Shose blvd., 1784 Sofia, Bulgaria }

\address[label2]{Armagh Observatory, College Hill, Armagh BT61 9DG, N. Ireland, UK}

\begin{abstract}

The present study provides important details on homologous eruptions of a solar prominence that occurred in active region NOAA 10904 on 2006 August 22.
We report  on the pre-eruptive phase of the homologous feature as well as  the kinematics and the morphology of a forth from a series of prominence eruptions that is critical in defining the nature of the previous consecutive eruptions.  The evolution of the overlying coronal field during homologous eruptions is discussed and a new observational criterion for homologous eruptions is provided.
We find a distinctive sequence of three activation periods  each of them containing  pre-eruptive precursors such as a brightening and enlarging of the prominence body followed by  small surge-like ejections from its southern end observed in the radio 17~GHz. We analyse a  fourth eruption that clearly indicates a full reformation of the prominence after the third eruption. The fourth eruption although occurring 11~hrs later has an identical morphology, the same angle of propagation with respect to the radial direction, as well as similar kinematic evolution as the previous three eruptions. We find an important feature of the homologous eruptive prominence sequence that is the maximum height increase of each consecutive eruption. 
The present analysis establishes that all four eruptions observed in H$\alpha$ are of confined  type with the third eruption undergoing a thermal disappearance during its eruptive phase. 
We suggest that the observation of the same direction of the magnetic flux rope (MFR) ejections can be consider as an additional observational criterion for MFR homology. This observational  indication for homologous eruptions  is  important, especially in the case of events of typical or poorly distinguishable morphology of eruptive solar phenomena.
\end{abstract}


\end{frontmatter}


\section{Introduction}
\label{Introduction}
The relationship between eruptive prominences (EPs) and other eruptive solar phenomena such as CMEs and  flares 
\citep[e.g.][]{1991SoPh..136..379S,2001ApJ...561..372S,2008ApJ...674..586S,2010SoPh..261..127C} suggests that the three 
eruptive events  often occur in the same large-scale coronal magnetic field configuration \citep[e.g.][]{2000JGR...10523153F}
 in which the EP occupies a limited volume  at its base. It is commonly accepted that 
solar prominence (filament) eruptions frequently accompany coronal mass ejections (CMEs). Thus, studying the pre-eruption phase, 
origin and evolution of EPs gives additional information relevant to CMEs' launch and propagation.

The observations and studies of early stages of prominence eruptions, i.e. prominence pre-eruptive activation, are crucial for the understanding of the signatures and pre-cursors of forthcoming solar eruptions. The observations of  prominence motions 
before and near the eruption onset can provide information for the coronal magnetic field evolution during the 
pre-eruptive stages \citep[e.g.][]{2012ApJ...761...69S}. Multi-wavelength studies 
of the precursor signatures for eruptions, such as pre-eruptive brightenings in microwave, extreme ultraviolet (EUV), and X-ray emission changes are  necessary to reveal the processes involved in the prominence destabilisation. In particular, microwave 
observations can show the full temporal and spatial prominence (filament) evolution, from early 
pre-eruptive stages to the end of eruption \citep[e.g.][]{2006PASJ...58...69G}. Moreover,  brightness temperature enhancements in  microwave observations at 17~GHz 
and 34~GHz  are a clear signature of 
heating in prominences \citep[e.g.][]{1999ApJ...510..466H, 2000ApJ...533..557H, 2006A&A...458..965C, 2013PASJ...65S..11G}.

Among the wide variety of solar eruptions there is a specific type of so-called ``sympathetic'' eruptions. Sympathetic 
solar eruptions are defined as consecutive eruptions that occur within a relatively short time interval either in one 
complex active region (AR) \citep[e.g.][]{2009ApJ...703..757L} or in different active regions located at large distances from each other 
\citep[e.g.][]{2007ApJ...664L.131Z}. 

In addition to sympathetic eruptive events, there also exist the so-called ``homologous'' eruptions. 
This term was first introduced by \citet{1938ZA.....16..276W} and \citet{1984AdSpR...4...11W} for solar flares,  and by 
\citet{2002ApJ...566L.117Z} for coronal mass ejections. The authors define 
flare-CME 
events as ``homologous'' when they have the same surface source, an identical shape and location (in the coronagraph field 
of view), and are associated with homologous X-ray or EUV activities. Recently, in terms of magnetic flux 
ropes, the homologous definition was also applied to all three eruptive events, CMEs, 
flares, and EPs  \citep{2013ApJ...778L..29L}. It includes three criteria:  the homologous flux ropes must originate from the same region within the same AR,  the endpoints of the homologous flux ropes have to be  anchored in the same location and   the morphologies of the homologous flux ropes have to resemble each other.

Solar surges are other eruptive phenomena that exhibit homologous behaviour \citep[e.g.][]{2012ApJ...760..101W}. They represent collimated plasma ejections along straight or slightly curved trajectories  \citep{1973SoPh...28...95R}. They have typical peak velocities of 100--300~$km~s^{-1}$ and maximum heights of 10--200~Mm \citep{2000SoPh..196...79S}. Their lifetime  is in the range of 10--20~min \citep{1973SoPh...28...95R, 2007A&A...469..331J} and they can reoccur during an hour or more \citep{1984SoPh...94..133S, 1995SoPh..156..245S}. Their origin and evolution are mostly associated with magnetic flux emergence and cancellation, as well as with flaring active regions. Surges often appear twisted and spiralled \citep[e.g.][for reviews]{1992PASJ...44L.173S, 1994ApJ...425..326S,1999ApJ...513L..75C, 2004ApJ...610.1136L, 2007A&A...469..331J, 2012ApJ...752...70U}. 

There are different approaches to modeling solar eruptions, ranging from 2D
analytical models to 3D numerical simulations \citep[][for reviews]{2000JGR...10523153F, 2009A&A...501..761M}.
Some of them use a breakout model \citep[e.g.][]{1999ApJ...510..485A, 2008ApJ...680..740D} to produce homologous eruptions.

In a recent study by \citet[][hereafter Paper I]{2014BlgAJ..21...92D} three  homologous prominence eruptions that occurred on 2006  August 22 in AR NOAA 10904 were examined. The consecutive eruptions were observed at the solar 
limb between 04:48 UT and 07:32 UT with the H$\alpha$ coronagraph at the National Astronomical Observatory Rozhen 
(NAO-Rozhen)  taken with a 1.8~\AA\ H$\alpha$ filter. The successive eruptions were associated with the same fragment from the AR filament. The kinematic 
patterns and evolutions of the first two eruptions classified them as confined. The third eruption, linked to 
a narrow CME and a type III radio burst at 164 MHz, was not fully understood because of the early prominence disappearance in H$\alpha$. 
The similar coronagraphic appearance of the eruptions and their non-radial propagations at approximately the same angle 
of $\approx50^\circ$ to the radial direction strongly suggests that the filament fragment underwent a triple homologous eruption.

The present research provides important findings on the pre-eruptive activity of homologous events by studying in great detail the 
pre-eruptive prominence activation in  radio images taken at frequency of 17 GHz (10\arcsec\ resolution and 10 min cadence) of the Nobeyama
Radioheliograph (NoRH). We report a detailed study  on a fourth consecutive homologous eruption that occurs 11~hrs later  using 
H$\alpha$ images (2.9\arcsec\ resolution and 3~min cadence) 
obtained by the Polarimeter for Inner Coronal Studies (PICS) instrument at the Mauna Loa Solar Observatory (MLSO), Hawaii. 
The rest of the data related to this event are given in Paper I.  The new additional analysis given here is crucial 
for the full understanding of  the evolution of the homologous sequence of  four prominence eruptions and the formulation of  an observational evidence for the definition of a solar eruption as homologous. We also provide important information on the evolution of the overlying coronal field during the homologous eruptions. The results are given in Section~2,  the pre-eruptive phase is reported in  Section~2.1 and the forth prominence eruption is described in Section~2.2. The discussion and conclusions are presented in Section~3.

\section{Results}
\label{Results}

 \subsection{Pre-activation phase}
\label{pre-act}

EPs, flares or CMEs often show pre-eruptive thermal or non-thermal signatures. Thermal signatures typically appear as a weak increase in the soft X-ray (SXR) light curve, while non-thermal are usually observed at radio wavelengths \citep{2006SSRv..123..303G}. 

\begin{figure}    
   \centerline{\includegraphics[width=0.92\textwidth,clip=]{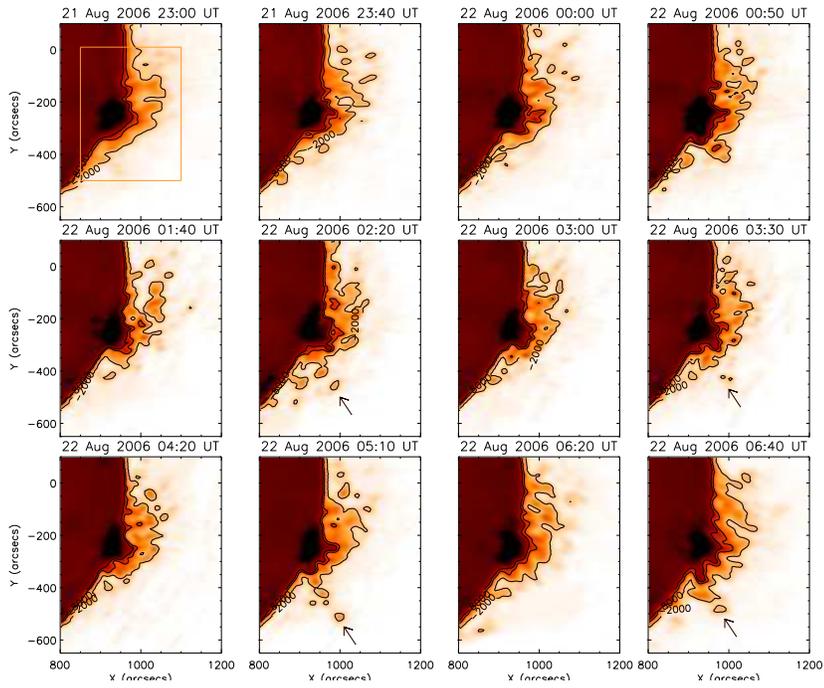}}
              \caption{Series of NoRH radio images at 17~GHz tracing the
pre-eruptive states,  the first and part of the second eruption of the prominence. The contours correspond to a brightness temperature  $T_b$ of 2000, 6000, and 8000~K.
                      }
   \label{fig1}
   \end{figure}

The sequence of prominence eruptions reported here was preceded by a distinctive pre-eruptive 
prominence activation that was observed in the NoRH radio 
data taken at a frequency of 17~GHz. These observations cover the quiet and 
pre-eruptive phases of the EP, the first, and part of the second eruptions (Fig.~\ref{fig1}).  We established a sequence of three activation periods  each of them containing pre-eruptive precursors such as a brightening and enlarging of the prominence body followed by small surge-like ejections.

   \begin{figure}    
   \centerline{\includegraphics[width=0.75\textwidth,clip=]{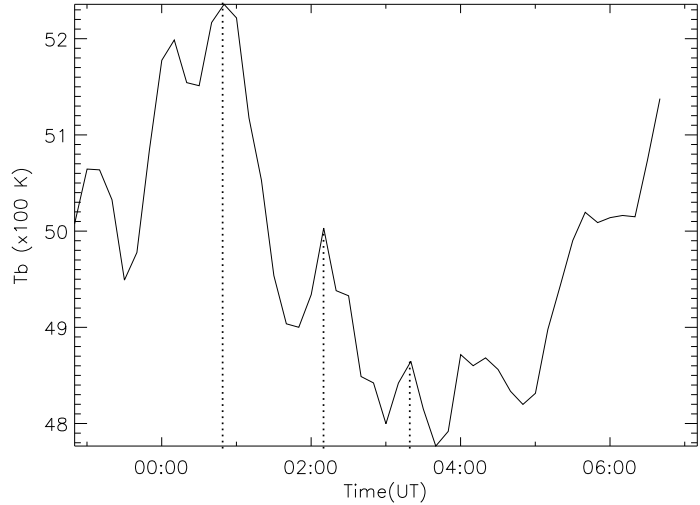}
              }
              \caption{Averaged radio flux at 17~GHz obtained from the boxed region overplotted on the first panel of Fig.~\ref{fig1}. The start time is 2006 August 21 at 22:50~UT. The dotted vertical lines indicate the emission peaks at 00:50~UT, 02:20~UT and 03:30~UT.
                      }
   \label{fig2}
   \end{figure}
   
   \begin{figure}    
   \centerline{\includegraphics[width=0.75\textwidth,clip=]{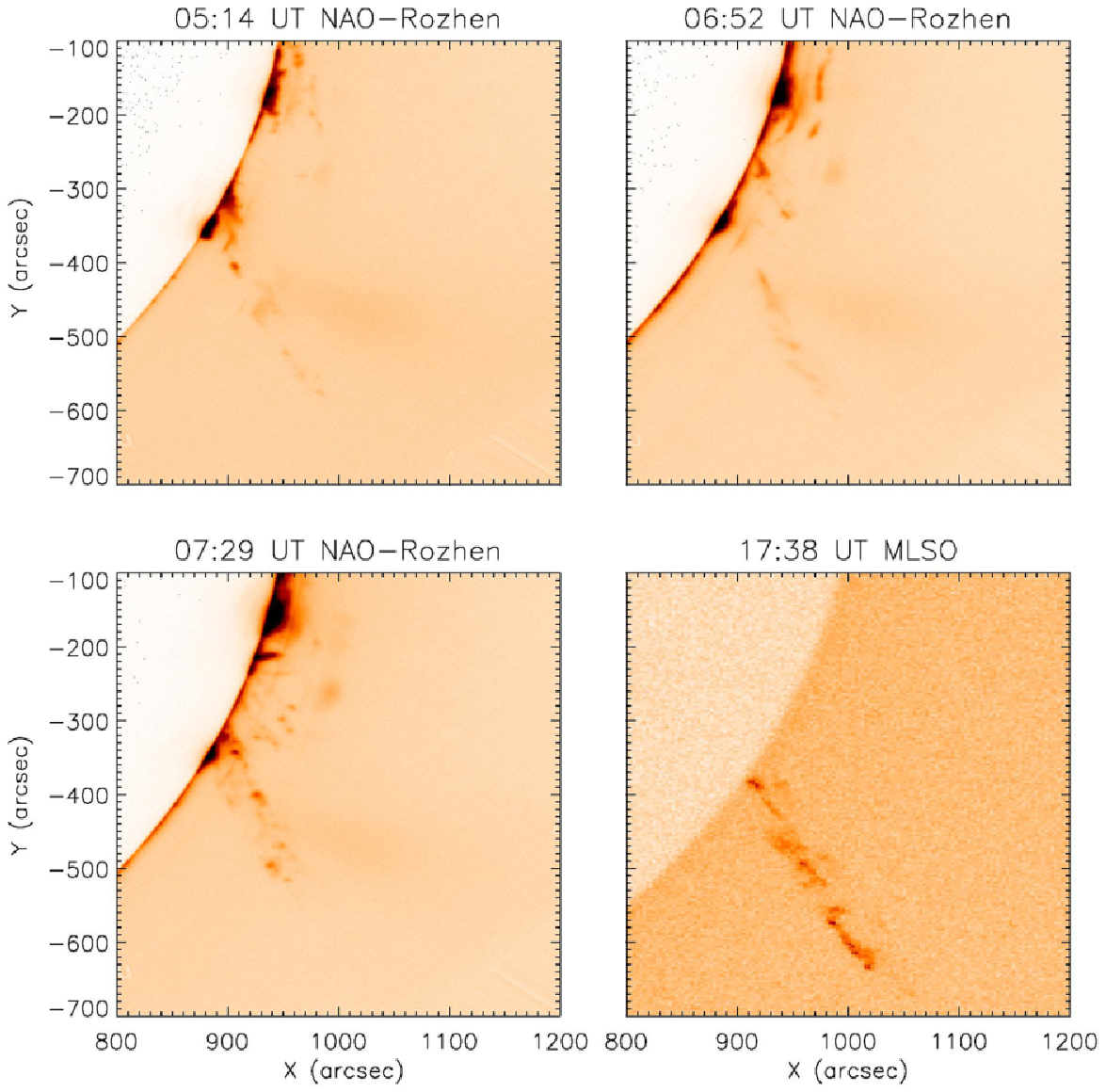}
              }
              \caption{Four prominence eruptions at their maximum height in the H$\alpha$ observations of NAO--Rozhen and MLSO.
                      }
   \label{fig3}
   \end{figure}
   
Until 00:00~UT on August 22, the prominence was in a  relatively  quiet phase. After this time the prominence body started changing 
its size, shape, and morphology, i.e. the first activation period began. The prominence body initially enlarged by 
stretching along the limb, which is most probably due to a heating of the prominence plasma \citep[e.g.][]{1999ApJ...510..466H}. 
This led subsequently to the increase of the brightness temperature Tb shown in Fig.~\ref{fig2}.
The increased Tb can be explained by a combination of optically thick emission from the cool prominence core and an 
optically thin emission from the heated prominence-corona transition region above $\sim10^{4}$~K \citep{2013PASJ...65S..11G}. The dynamic  evolution of the body shape and morphology  can be followed in the online material (Fig.~\ref{fig1o}). These changes suggest a heating and a turbulence increase, which have led to the prominence fragmentation. Between approximately  00:00~UT and
01:40~UT, in addition to the aforementioned pre-eruptive events, at 
the southern end of the prominence body a small surge-like ejection appeared at 00:50~UT (Fig.~\ref{fig1}) that
lasted until  01:10~UT (see the online material).

The described sequence of pre-eruptive events is repeated again between 01:40~UT and 03:00~UT (second activation period) and later between 
03:00~UT and 04:20~UT (third activation period). Each period is characterized by small surge-like ejections that occurred in the same place 
of the prominence body (Fig.~\ref{fig1}), where later all four prominence eruptions were observed  (Fig.~\ref{fig3}). Moreover, they are associated with  peaks in the radio flux at 17~GHz marked with dotted lines in Fig.~\ref{fig2}.

\begin{figure}    
   \centerline{\includegraphics[width=0.9\textwidth,clip=]{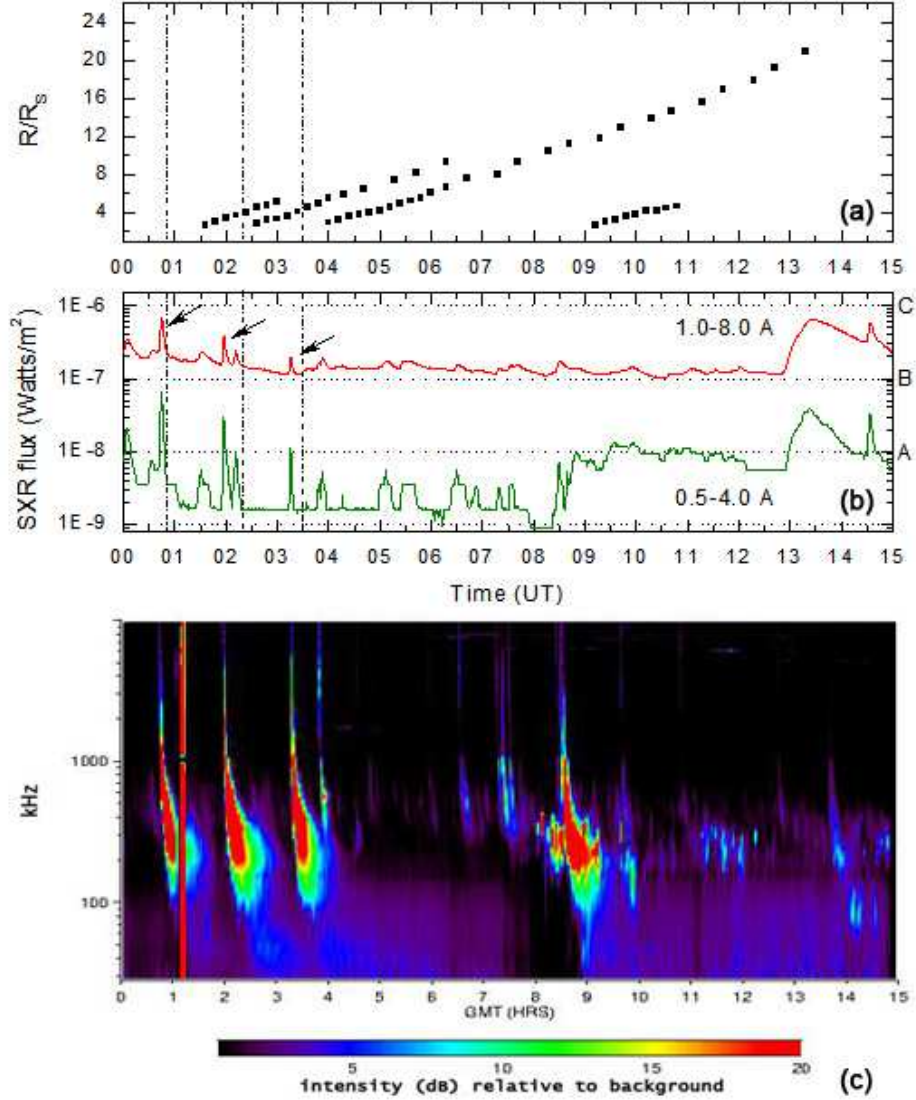}
              }
              \caption{(\textbf{a}) Height-time profiles of the CMEs associated with AR 10904. (\textbf{b}) GOES-12 1-min SXR fluxes in the 0.5-4.0\AA\ and 1.0-8.0~\AA\ channels. (\textbf{c}) WIND/WAVES RAD-1 and RAD-2 dynamic spectrum. The vertical dash-dotted lines in the panels \textbf{a} and \textbf{b} mark the times of the maximum phase of three pre-eruptive periods during the prominence activation. The three black arrows in panel \textbf{b} point at the B-class flares linked to the pre-eruptive phases.
                      }
   \label{fig4}
   \end{figure}

Each of these three pre-eruptive events was linked to weak narrow CMEs according to the LASCO CME catalogue (Fig.~\ref{fig4}a). Weak three  B-class SXR flares with a source AR 10904 (Fig.~\ref{fig4}b) and three consecutive type III bursts (WIND/WAVES, Fig.~\ref{fig4}c) were also recorded. The association of  a prominence/filament pre-eruptive activity with slow and narrow CMEs, and weak SXR flares is  typical for ARs with $\beta$ magnetic field configuration  \citep{2011MNRAS.414.2803Y} which is also the case for AR 10904 (see Paper~I).

  \subsection{Fourth prominence eruption}
\label{Fourth eruption}

\begin{figure}    
   \centerline{\includegraphics[width=0.82\textwidth,clip=]{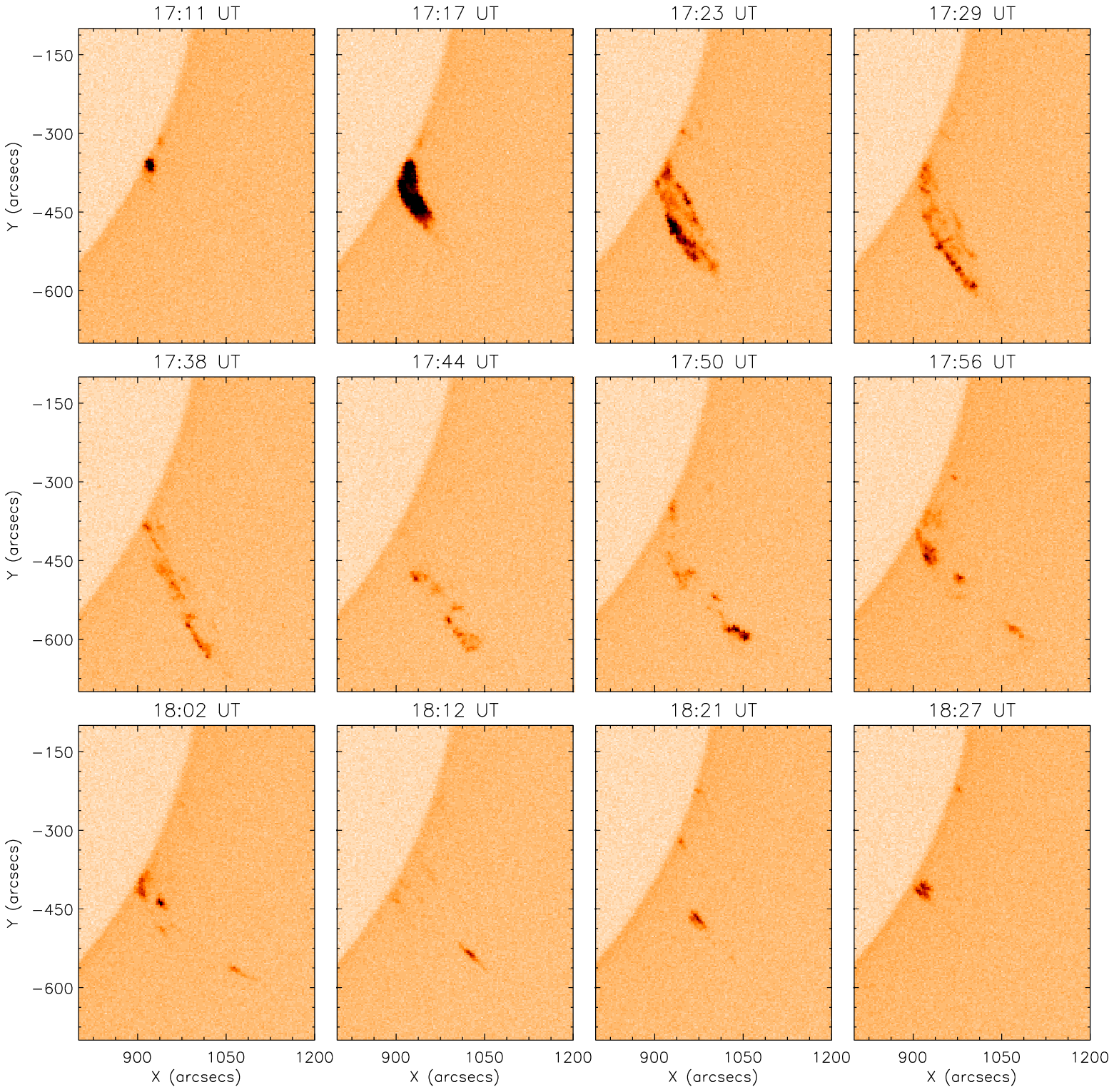}
              }
              \caption{Series of MLSO $H\alpha$ images tracing the fourth prominence eruption
on 2006 August 22. 
                      }
   \label{fig5}
   \end{figure}

\begin{figure}    
   \centerline{\includegraphics[width=0.75\textwidth,clip=]{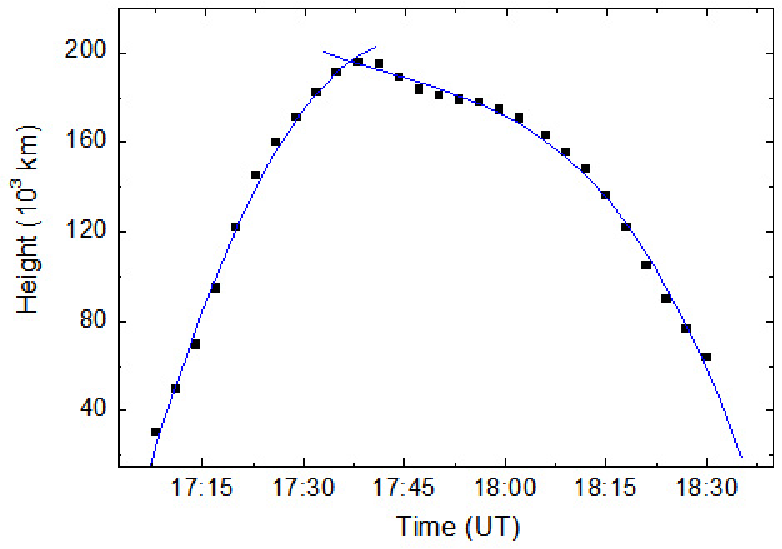}
              }
              \caption{Height-time profile of the fourth prominence eruption.
                      }
   \label{fig6}
   \end{figure}

We found that a fourth eruption registered by MLSO in $H\alpha$ occurred on August 22 starting $\sim$11 hours after the third eruption.
It first appeared in the MLSO PICS FOV at 17:08~UT and half an hour 
later, at 17:38~UT, it reached a maximum height of 196~Mm. After that time, the prominence 
plasma started to flow back to the chromosphere and after 18:30~UT the prominence completely 
disappeared behind the PICS occulting disk (Fig.~\ref{fig5}).  The prominence eruption 
lasted 82 minutes  until 18:30~UT.
Its height-time profile,  given in Fig.~\ref{fig6}, is similar to those of the first and second eruptions shown in 
Figs.~\ref{fig2}  and \ref{fig3} of Paper~I. The fourth eruption does not show initial 
acceleration. Possibly,  the prominence did undergo acceleration but that was not observed for two reasons. First, when the eruption took 
place 11 hours later, the filament fragment was already about $9^\circ$ behind 
the limb. The second reason is purely instrumental as the diameter 
of the MLSO PICS occulting disk is larger than  the solar disk with 73.06~arcsec, i.e. the 
occulting disk exceeds the solar limb with 36.53~arcsec ($\approx$26.5~Mm). Moreover, as it
can be seen in Figs.~2 and 3 in Paper I, the acceleration phases of the first two prominence eruptions reached a height of up to 50~Mm.

In order to evaluate the kinematic patterns of the eruptive phase of the fourth eruption, a 2nd-order 
polynomial fit was used. The fit suggests that the prominence rose with a negative 
acceleration of $-$72~$m~s^{-2}$. The eruption velocity decreased from 160~$km~s^{-1}$  at 17:08~UT up to 
6~$km~s^{-1}$ at 17:58~UT when the prominence reached a maximum height of 196~Mm.
The downflow phase was estimated by a 3rd-order polynomial fit. The polynomial model 
gives a plasma downflow deceleration, which initially changed from $-$48.3~$m~s^{-2}$ to 
$-$3.3~$m~s^{-2}$ at 18:12~UT (149~Mm) and then increased up to 23.7~$m~s^{-2}$. The velocity 
decreased from 62~$km~s^{-1}$ to 15~$km~s^{-1}$ at 18:12~UT and subsequently increased to 26~$km~s^{-1}$. 
An important feature of the downflow kinematics  is the change of  the acceleration sign from negative 
to positive between 18:12~UT and 18:15~UT. This evolution (e.g. untwist of the MFR) is crucial for the conditions at which one of the two forces  (magnetic tension and gravity force) acting on the EP plasma will become predominant. As a consequence a change of the acceleration will occur.  As it can be seen in Fig.~\ref{fig5} and the online animation (Fig.~\ref{fig2o}), the angle between the EP MFR and  the radial direction through its base at the limb started to decrease after 17:56~UT. This process was accompanied by the untwisting of the MFR until 18:09~UT. At 18:15~UT  the threads of the MFR were completely untwisted. Therefore, before 18:12~UT the main factor for the plasma downflow was the holding  up action of the magnetic tension in  a tightly twisted MFR, which strongly dominated over the 
gravity force. After 18:15~UT the gravity force was the main driving mechanism for the plasma downflow. 

The comparison of the fourth eruption with previous ones (Paper~I) shows that all four prominence eruptions preserve their non-radial 
propagation upwards in the solar corona, i.e. the direction of propagation is identical for all four eruptions. The fourth eruption, as three previous ones, started at the same point at the solar limb, propagated under approximately the same angle of $\approx50^\circ$ to the radial direction, and had similar helical morphology.
This suggests an almost complete restoration of the magnetic skeleton between the successive eruptions. 
The prominence appeared  more compact in the MLSO images, which is due to the broader band $H\alpha$ filter and the lower spatial resolution.

The fourth prominence eruption, similar to the first and second,  has  a height-time profile that comply with a confined type of  
eruption. The arguments for this assumption are: first, we clearly observe a 
downflow process during these eruptions without the detection of a breaking of the MFR; second, except for the third eruption, no CMEs were  associated  with the other three eruptions. The identification and analysis of  the fourth eruption in a homologous sequence allowed us  to define the type of the third eruption. The gradual disappearance of the prominence during the  third eruption in $H\alpha$ is caused by a heating process which is typical for the so-called ``thermal disparition brusque'' (DB) of solar filaments \citep{1989HvaOB..13..379M}. Despite the fact that this eruption was  associated with a narrow CME, it is now clear that it also represents a confined (failed) eruption. The observation of a fourth eruption, which has similar morphology and dynamics as well as the direction of propagation supports this conclusion. 

 \section{Discussion and Conclusions}
\label{Discussion}

The importance of homologous event studies consists in the understanding of how magnetic energy is stored and released in the solar atmosphere, and how the magnetic field is reformed. The present study shows 
that the event on 2006 August 22 represents a sequence of four prominence eruptions with the fourth eruption taking place 11 hrs later than the previous three eruptions. Each eruption has an almost identical  appearance in the shape of a  surge-like non-radial ejection of an EP MFR. We established that all four eruptions clearly represent a homologous event according to the criteria of \citet{2013ApJ...778L..29L}. 

Although, we define the three events   recorded in the radio 17~GHz images  between 00:00~UT and 04:20~UT on August 22 as pre-activation phase  of the eruptive events observed in H$\alpha$, we cannot dismiss the possibility that these ejections  were also parts of the homologous eruptive sequence reported here. Please note that the radio eruptions also  display a typical  homologous behaviour. Unfortunately, the lack of H$\alpha$ and EUV observations during this period of time prevents us to firmly confirm this.

The prominence eruptions studied here show certain similarities with surge events. Their outward and downward  motions take place along the same nearly straight trajectory, and the height and  velocity ranges of the eruptions are similar to those of surges. In spite of that we believe that the observed homologous prominence eruptions seem to differ from typical surge events based on the following arguments:

(1) Observationally, surges are always found in newly emerging active regions  \citep{1973SoPh...28...95R} and are often closely associated with flares.  Surges usually either shortly precede or follow flares (with $\sim$5~min) \citep[e.g.][]{1988A&A...201..327S, 1995SoPh..156..245S, 2012ApJ...752...70U}.  In contrast, the homologous eruptions analysed here occurred during the decay phase  of the active region.

(2) The duration of the homologous prominence eruptions are between 54 and 82~min (see Section~2.2), which is almost 2--3 times longer than the typical 10--20~min duration of surges \citep[e.g.][]{1973SoPh...28...95R, 2007A&A...469..331J}.

(3) The four homologous EPs were preceded by a  distinctive pre-eruptive prominence activation (Section~2.1)  during $\sim$5 hours while surges do not present such phases.

With regard to the differences between surges and EPs given above, we believe that the phenomena studied here represent homologous prominence rather than surge eruptions.
It is important to note that because the footpoints of the eruptive prominence are located 9$^\circ$ behind the limb, we cannot, therefore, judge whether flux emergence has taken place at this location.

An important feature of the studied events is that the source of the four
homologous eruptions is positioned at the north edge of the helmet streamer as mentioned in Paper~I,
i.e. it is highly asymmetrical with respect to  the overlying arcades of the streamer.
A coronal helmet streamer has a strong influence on the early propagation of an EP \citep[e.g.][]{2011NewA...16..276B}. 
As suggested by \citet{2007ApJ...661..543M}, the magnetic field in the guiding streamer leg needs to have the necessary strength to laterally
channel the EP to the nearby arcade. In the present case, the same non-radially directed motion of all eruptions clearly demonstrate  this guiding action.
The confinement of the homologous prominence eruptions can be explained 
with the asymmetrical position of the filament inside the helmet streamer. Such an asymmetry of the 
background fields is one of the crucial factors for confined eruptions \citep{2009ApJ...696L..70L}.

In the present study, each consecutive prominence eruption, except the third one that presents signatures of  thermal disappearance (e.g. its disappearance in H$\alpha$ and association with a metric type III radio burst (Paper I)), reached bigger height (Fig.~\ref{fig3}). A physical mechanism for the height increase during  successful prominence eruptions is given by  \citet{2013ApJ...769L..25C}, who suggest that each  eruption partially opens (weakens) the overlying magnetic field and thus decreases its magnetic restriction.
The occurrence of the fourth eruption at the same place 11 hours 
later and its similarity to the previous eruptions, including the helical morphology and the non-radial propagation, suggests
a filament reformation at the same place after the third eruption. An additional argument for the filament 
reformation is the gradual EP disappearance in H$\alpha$ suggesting  a thermal disappearance that is always followed by a reappearance after the prominence/filament cools down to the H$\alpha$ formation temperature  \citep{1989HvaOB..13..379M}.

It is important to note that the multiple non-radial homologous prominence 
eruptions observed at the solar limb are rarely reported phenomena (although they are possibly not a rare solar phenomenon). The nearest similar 
case are six homologous filament eruptions in  active region NOAA 11045 on 2010 February 2008 studied by
\citet{2011RAA....11..594S}. 

\section*{Acknowledgments}

We are grateful to the instrumental teams of the Nobeyama Radioheliograph at the National Astronomical Observatory of
Japan for their open-data policy. Courtesy of the Mauna Loa Solar Observatory, operated by the High Altitude Observatory, as part of the National Center for Atmospheric Research (NCAR). NCAR is supported by the National Science Foundation.  MM is supported by the Leverhulme Trust.
The Wind/WAVES instrument is a joint effort of the Paris-Meudon Observatory, the University of Minnesota, and the Goddard Space Flight Center. We thank the GOES team for data support.

\section*{References}
 \bibliographystyle{elsarticle-harv} 
 \bibliography{ref1.bib}







 
  \appendix
 \label{appendix}
\section{Online material}

  \begin{figure}[!h]    
   \centerline{\includegraphics[width=0.4\textwidth,clip=]{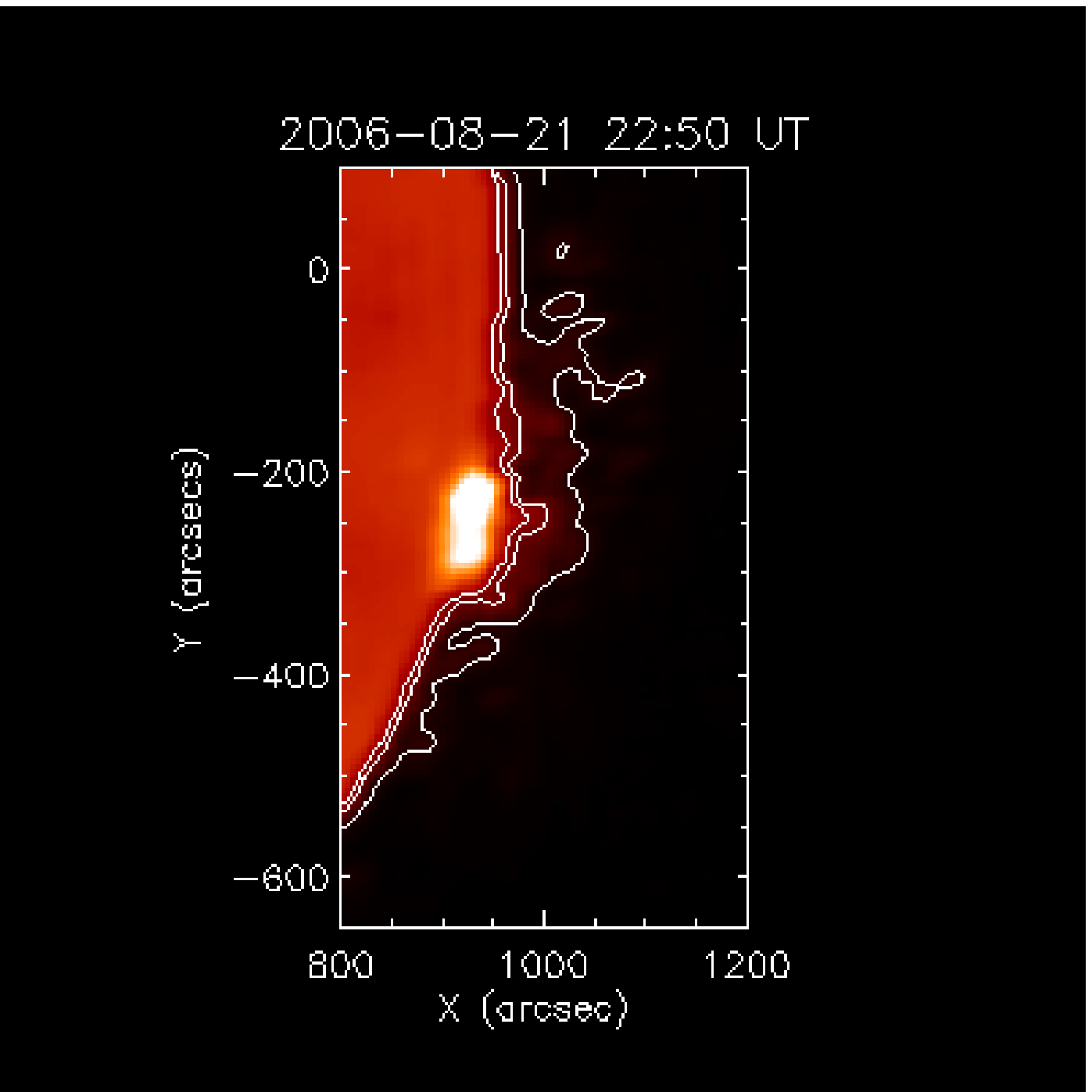}
              }
   \caption{Animation of the 17~GHz images.  The contours correspond to a brightness temperature  $T_b$ of 2000, 6000, and 8000~K.}
    \label{fig1o}
  \end{figure}

  \begin{figure}[!h]    
   \centerline{\includegraphics[width=0.4\textwidth,clip=]{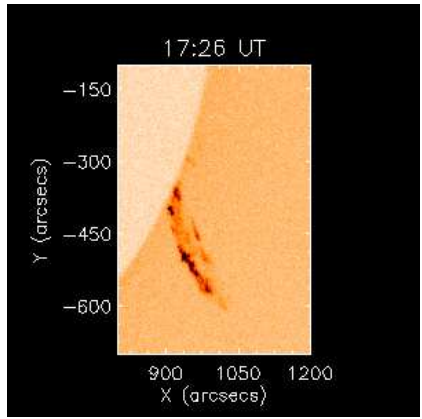}
              }
   \caption{Animation  of MLSO $H\alpha$ images of  the fourth prominence eruption on 2006 August 22.}
    \label{fig2o}
  \end{figure}
   
\end{document}